\documentclass[10pt,conference]{IEEEtran}
\IEEEoverridecommandlockouts
\usepackage{cite}
\usepackage{amsmath,amssymb,amsfonts}
\usepackage{algorithmic}
\usepackage{multirow}
\usepackage{array}
\usepackage{graphicx}
\usepackage{textcomp}
\usepackage{xcolor}
\usepackage{booktabs}
\usepackage{tabularx}
\usepackage{subcaption}
\usepackage{hyperref}

\def\BibTeX{{\rm B\kern-.05em{\sc i\kern-.025em b}\kern-.08em
    T\kern-.1667em\lower.7ex\hbox{E}\kern-.125emX}}

\begin{document}

\title{Combining Large Language Models with Static Analyzers for Code Review Generation
\thanks{The replication package is available at \url{https://github.com/ImenJaoua/Hybrid-Code-Review} and the data is available at \url{https://zenodo.org/records/14061110}.}
}

\author{\IEEEauthorblockN{Imen Jaoua}
\IEEEauthorblockA{\textit{Université de Montréal} \\
Montreal, Canada \\
imen.jaoua@umontreal.ca}
\and
\IEEEauthorblockN{Oussama Ben Sghaier}
\IEEEauthorblockA{\textit{Université de Montréal} \\
Montreal, Canada \\
oussama.ben.sghaier@umontreal.ca}
\and
\IEEEauthorblockN{Houari Sahraoui}
\IEEEauthorblockA{\textit{Université de Montréal} \\
Montreal, Canada \\
sahraouh@iro.umontreal.ca}

}

\maketitle

\begin{abstract}
Code review is a crucial but often complex, subjective, and time-consuming activity in software development. Over the past decades, significant efforts have been made to automate this process. Early approaches focused on knowledge-based systems (KBS) that apply rule-based mechanisms to detect code issues, providing precise feedback but struggling with complex, context-dependent cases. More recent work has shifted toward fine-tuning pre-trained language models for code review, enabling broader issue coverage but often at the expense of precision. In this paper, we propose a hybrid approach that combines the strengths of KBS and learning-based systems (LBS) to generate high-quality, comprehensive code reviews. Our method integrates knowledge at three distinct stages of the language model pipeline: during data preparation (Data-Augmented Training, DAT), at inference (Retrieval-Augmented Generation, RAG), and after inference (Naive Concatenation of Outputs, NCO). We empirically evaluate our combination strategies against standalone KBS and LBS fine-tuned on a real-world dataset. Our results show that these hybrid strategies enhance the relevance, completeness, and overall quality of review comments, effectively bridging the gap between rule-based tools and deep learning models.


\end{abstract}

\section{Introduction}
\label{sec:intro}

Code review is a common software engineering practice that plays a crucial role in maintaining code quality, identifying bugs, and fostering a culture of continuous improvement within development teams \cite{ccetin2021review,xiaomeng2018survey}. By rigorously inspecting code changes, reviewers provide valuable feedback that can improve software both before and after it is integrated into the system. Modern code review goes beyond defect detection; it aims to enhance the overall quality of software changes by ensuring maintainability, reliability, and adherence to best practices \cite{hong2022commentfinder, bavota2015four}.

However, the code review process is often perceived as complex, time-consuming, and subject to various biases, especially in large-scale projects. Factors such as varying levels of developer expertise, interpersonal dynamics, and the lack of standardized guidelines can introduce inconsistencies, ultimately impacting the efficiency and robustness of the codebase~\cite{ben2024improving}. Furthermore, the evolving nature of coding standards and best practices requires constant adaptation, adding further complexity to the process.

To address these challenges, there has been growing interest in automating the code review process \cite{hovemeyer2004finding}. Initial efforts primarily involved the deployment of knowledge-based systems (KBS), particularly static analysis tools like PMD, FindBugs, and SonarQube. These tools use predefined rule sets to identify common coding issues, systematically scanning the codebase for violations and improving the early stages of development by flagging rule violations in the code. Although effective at detecting surface-level issues, static analyzers are limited by their dependence on manually defined rules, which require frequent updates to remain relevant. Their rigidity makes it difficult to handle complex, context-dependent issues that require a deeper understanding of code intent \cite{sadowski2015tricorder}. Consequently, these tools often fall short in adapting to the dynamic and context-sensitive nature of software projects, where factors like architecture, team culture, and specific project requirements demand more flexible and intelligent solutions.

Recent advances in large language models and natural language processing have sparked significant interest in using pre-trained language models to automate code review workflows. These learning-based systems (LBS) aim to overcome the limitations of knowledge-based detection by providing a more nuanced understanding of code, enabling them to identify deeper faults, recommend improvements, and even anticipate the potential impact of code changes on the overall system~\cite{wadhwa2024core}. Although these models capture a broader range of issue patterns than static analysis alone, their precision still remains below acceptable levels \cite{tufano2024code, ibtasham2024towards}.

As is common in the automation of software development tasks, some solutions achieve high precision but with limited coverage, while others offer broader coverage at the expense of precision. The most promising approaches, therefore, are those that balance these trade-offs by combining solutions to optimize both precision and coverage. We hypothesize that integrating static analysis with learning-based systems can enhance the effectiveness of automated code review generation tools.
In this paper, we explore three combination strategies aimed at integrating the structured knowledge from static analysis into the pipeline for building and operating a large language model (LLM) for code review generation. Specifically, our approach integrates knowledge at three distinct points in the pipeline: during data preparation for fine-tuning (data-augmented training, DAT), at inference time (retrieval-augmented generation, RAG), and after inference (naive concatenation of outputs, NCO).

To evaluate these strategies, we used a combination of human judgments and LLM-based assessments to compare the generated reviews from each approach with baseline systems: standalone KBS and LBS. Our results show that combining KBS and LBS captures the strengths of both: the precision of static analysis and the comprehensiveness of LLMs, resulting in more effective and human-like code reviews.

The remainder of this paper is organized as follows: Section~\ref{sec:background} provides an overview of the background. Section~\ref{sec:approach} details the components of our proposed approach. Section~\ref{sec:evaluation} presents the evaluation results. Section~\ref{sec:literature} discusses related work. Finally, Section~\ref{sec:conclusion} offers concluding remarks and suggests potential directions for future improvements to our approach.

\section{Background}
\label{sec:background}

\subsection{Code Review Automation}
Code review is a widely adopted practice among software developers where a reviewer examines changes submitted in a pull request \cite{hong2022commentfinder, ben2024improving, siow2020core}. If the pull request is not approved, the reviewer must describe the issues or improvements required, providing constructive feedback and identifying potential issues. This step involves review commment generation, which play a key role in the review process by generating review comments for a given code difference. These comments can be descriptive, offering detailed explanations of the issues, or actionable, suggesting specific solutions to address the problems identified \cite{ben2024improving}.

Various approaches have been explored to automate the code review comments process  \cite{tufano2023automating, tufano2024code, yang2024survey}. 
Early efforts centered on knowledge-based systems, which are designed to detect common issues in code. Although these traditional tools provide some support to programmers, they often fall short in addressing complex scenarios encountered during code reviews \cite{dehaerne2022code}. More recently, with advancements in deep learning, researchers have shifted their focus toward using large-language models to enhance the effectiveness of code issue detection and code review comment generation.

\subsection{Knowledge-based Code Review Comments Automation}

Knowledge-based systems (KBS) are software applications designed to emulate human expertise in specific domains by using a collection of rules, logic, and expert knowledge. KBS often consist of facts, rules, an explanation facility, and knowledge acquisition. In the context of software development, these systems are used to analyze the source code, identifying issues such as coding standard violations, bugs, and inefficiencies~\cite{singh2017evaluating, delaitre2015evaluating, ayewah2008using, habchi2018adopting}. By applying a vast set of predefined rules and best practices, they provide automated feedback and recommendations to developers. Tools such as FindBugs \cite{findBugs}, PMD \cite{pmd}, Checkstyle \cite{checkstyle}, and SonarQube \cite{sonarqube} are prominent examples of knowledge-based systems in code analysis, often referred to as static analyzers. These tools have been utilized since the early 1960s, initially to optimize compiler operations, and have since expanded to include debugging tools and software development frameworks \cite{stefanovic2020static, beller2016analyzing}.

\subsection{LLMs-based Code Review Comments Automation}
As the field of machine learning in software engineering evolves, researchers are increasingly leveraging machine learning (ML) and deep learning (DL) techniques to automate the generation of review comments \cite{li2022automating, tufano2022using, balachandran2013reducing, siow2020core, li2022auger, hong2022commentfinder}. Large language models (LLMs) are large-scale Transformer models, which are distinguished by their large number of parameters and extensive pre-training on diverse datasets.  Recently, LLMs have made substantial progress and have been applied across a broad spectrum of domains. Within the software engineering field, LLMs can be categorized into two main types: unified language models and code-specific models, each serving distinct purposes \cite{lu2023llama}.

Code-specific LLMs, such as CodeGen \cite{nijkamp2022codegen}, StarCoder \cite{li2023starcoder} and CodeLlama \cite{roziere2023code} are optimized to excel in code comprehension, code generation, and other programming-related tasks. These specialized models are increasingly utilized in code review activities to detect potential issues, suggest improvements, and automate review comments \cite{yang2024survey, lu2023llama}.

\subsection{Retrieval-Augmented Generation}
Retrieval-Augmented Generation (RAG) is a general paradigm that enhances LLMs outputs by including relevant information retrieved from external databases into the input prompt \cite{gao2023retrieval}. Traditional LLMs generate responses based solely on the extensive data used in pre-training, which can result in limitations, especially when it comes to domain-specific, time-sensitive, or highly specialized information. RAG addresses these limitations by dynamically retrieving pertinent external knowledge, expanding the model's informational scope and allowing it to generate responses that are more accurate, up-to-date, and contextually relevant \cite{arslan2024business}. 

To build an effective end-to-end RAG pipeline, the system must first establish a comprehensive knowledge base. It requires a retrieval model that captures the semantic meaning of presented data, ensuring relevant information is retrieved. Finally, a capable LLM integrates this retrieved knowledge to generate accurate and coherent results \cite{ibtasham2024towards}.

\subsection{LLM as a Judge Mechanism}

LLM evaluators, often referred to as LLM-as-a-Judge, have gained significant attention due to their ability to align closely with human evaluators' judgments \cite{zhu2023judgelm, shi2024judging}. Their adaptability and scalability make them highly suitable for handling an increasing volume of evaluative tasks. 

Recent studies have shown that certain LLMs, such as Llama-3 70B and GPT-4 Turbo, exhibit strong alignment with human evaluators, making them promising candidates for automated judgment tasks \cite{thakur2024judging}

To enable such evaluations, a proper benchmarking system should be set up with specific components: \emph{prompt design}, which clearly instructs the LLM to evaluate based on a given metric, such as accuracy, relevance, or coherence; \emph{response presentation}, guiding the LLM to present its verdicts in a structured format; and \emph{scoring}, enabling the LLM to assign a score according to a predefined scale \cite{ibtasham2024towards}. Additionally, this evaluation system can be enriched with the ability to explain reasoning behind verdicts, which is a significant advantage of LLM-based evaluation \cite{zheng2023judging}. The LLM can outline the criteria it used to reach its judgment, offering deeper insights into its decision-making process.

\section{Proposed approach}
\label{sec:approach}

For the task of review comment generation, Knowledge-Based Systems (KBS) draw on codified rules and expert knowledge to deliver feedback that is consistent with established coding standards and best practices. Static analyzers, a prominent example of KBS, systematically follow predefined guidelines to detect code issues, offering reliable and structured feedback. While KBS achieve high precision, they are limited in scope, covering only a subset of possible issues encountered during code changes. In contrast, Learning-Based Systems (LBS) harness the adaptive potential of language models, which, by training on historical data, can recognize intricate patterns and generate contextually relevant review comments. This adaptability allows LBS to cover a broader range of issues present in the dataset, though often at the expense of precision. In this work, we conjecture that by combining these two strategies, it is possible to achieve the best of both approaches, namely, broader issue coverage coupled with improved precision.

\subsection{Overview}

\begin{figure}[htbp!]
  \centering
  \includegraphics[width=1\linewidth]{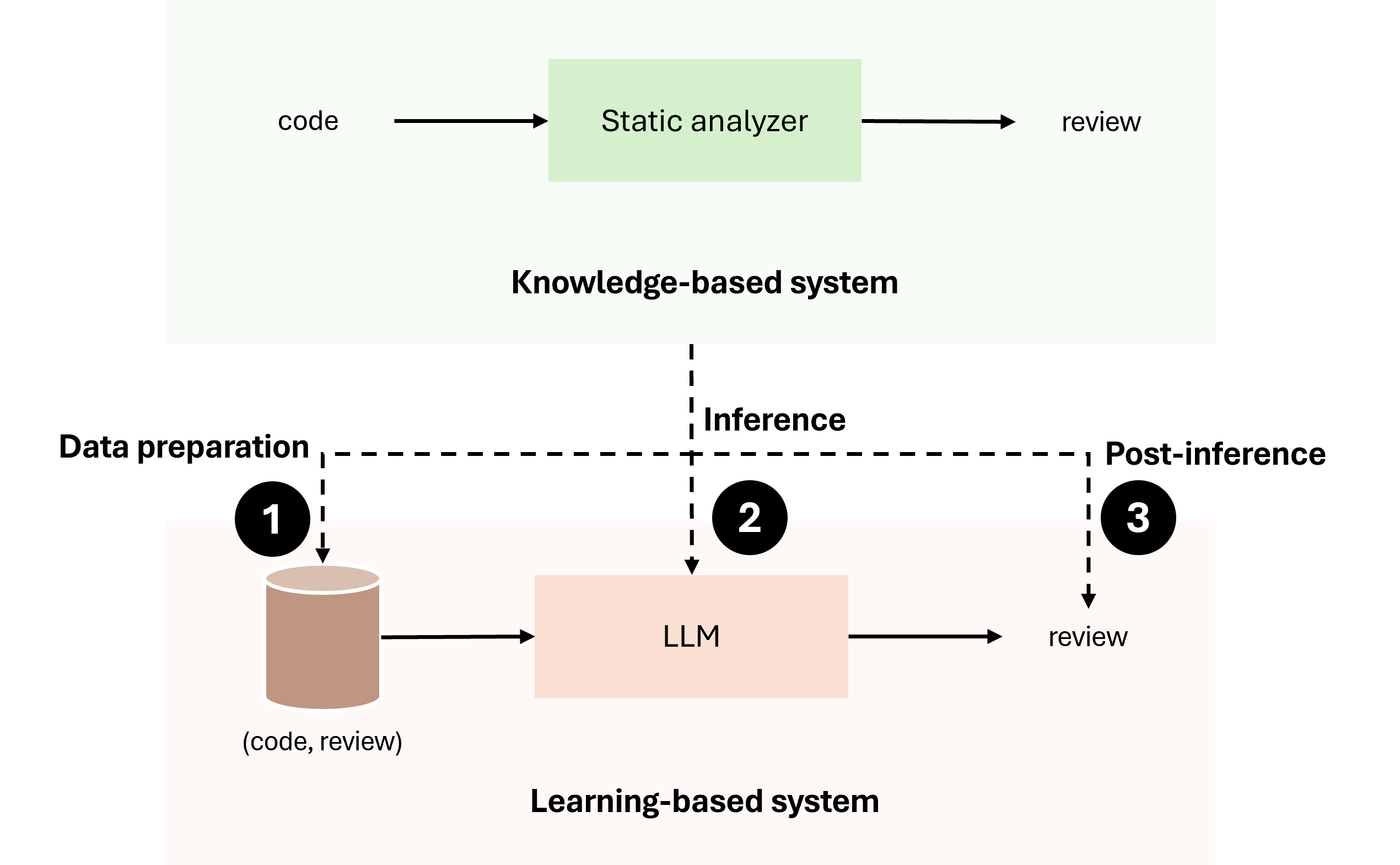}
  \caption{Different strategies to combine learning and knowledge-based systems}
  \label{fig:combination} 
\end{figure}

Figure~\ref{fig:combination} illustrates our approach, outlining three strategies to combine Knowledge-Based Systems (KBS) and Learning-Based Systems (LBS) to enhance code review automation. 

These strategies leverage KBS insights at different stages of the LBS pipeline, specifically during data preparation, inference, and final output. The three strategies are as follows:

\begin{itemize} 
    \item \emph{Data-Augmented Training (DAT)}: In this strategy, we enhance the training dataset by augmenting a real-world dataset with synthetic data generated from both KBS and LBS. This enriched dataset is then used to fine-tune a language model, enabling the LBS to incorporate both data-driven patterns and rule-based knowledge. This combination helps the model gain a more comprehensive understanding of code review patterns, improving robustness in varied review scenarios.
    \item \emph{Retrieval-Augmented Generation (RAG)}: Here, KBS insights are integrated directly into the LBS inference process. Through RAG, relevant information is dynamically retrieved from KBS (i.e., static analysis results) and injected into the prompts during generation. By incorporating the results of the KBS into the instruction, the LBS aligns its responses with established coding standards and practices, providing feedback grounded in structured, rule-based knowledge.
    \item \emph{Naive Concatenation of Outputs (NCO)}: This strategy merges the feedback generated by KBS and LBS after inference, combining their outputs to produce a unified code review. By consolidating KBS’s rule-based precision with LBS’s contextual depth, NCO offers a comprehensive review that covers a broader range of potential issues.
\end{itemize}

These strategies allow the LBS to benefit from the structured, rule-based insights of KBS, enhancing its ability to generate accurate, contextually appropriate, and standards-compliant code review comments.

\subsection{Baseline Model Preparation and Static Analyzers Selection}

While our approach is applicable to a wide range of LLMs and static analysis tools, we propose a specific configuration to illustrate the three strategies and establish the baselines for validating our conjecture. To set up the baseline systems, we first defined the LBS. We fine-tuned a large language model on an extensive code review dataset \cite{li2022automating}, referred to as \(\mathcal{D}_{\mathcal{M}_i}\), which pairs code changes with detailed reviews.

The selected model for fine-tuning is \emph{CodeLlama-7b}, trained for comment generation (i.e., generating review comments from code changes) with the following hyperparameter settings. The training was conducted on four \emph{NVIDIA RTX 3090} GPUs, using a batch size of $4$ per device. To boost efficiency, we applied gradient accumulation with a step size of $4$, updating the optimizer only after multiple batches. We used 4-bit quantization to improve memory and computational efficiency. Additionally, we employed Quantized Low-Rank Adaptation (QLoRA) \cite{hu2021lora}, a Parameter-Efficient Fine-Tuning (PEFT) technique, with $r = 16$, $\alpha = 32$, and $dropout = 0.05$. This method decomposes weight updates into low-rank matrices, reducing the parameters needed for fine-tuning and optimizing training efficiency \cite{hu2021lora}.

This resulted in a model, denoted as \(\mathcal{M}_i\), capable of generating detailed human-like code reviews.
Since \(\mathcal{M}_i\) represents the LBS component and was trained using the data from \(\mathcal{D}_{\mathcal{M}_i}\), we used the test set to generate reviews by both static analyzers and the fine-tuned model \(\mathcal{M}_i\).

To focus on a relevant subset of available static analyzers, we filtered the test set to include only Java code samples, producing a subset of $27,267$ entries, denoted \(\mathcal{D}_o\). Each entry in \(\mathcal{D}_o\) is a tuple \( (f,c) \), where \( f \) represents the source code file and \( c \) denotes the code change. Here, \( c \) is input to the LBS, while \( f \) serves as input to the KBS.
We limited our selection of static analyzers to tools that process Java source code directly. Although this decision excludes tools designed for Java bytecode analysis, it allows for a broader range of issue types. Specifically, we selected two well-established static analyzers: PMD \cite{pmd} and Checkstyle \cite{checkstyle}, both of which are designed to identify potential issues directly in source code.

PMD is a static code analysis tool that identifies issues in code by applying a set of rules aimed at detecting common problems, which are categorized into eight groups: best practices, coding style, design, documentation, error-prone, multi-threading, performance, and security \cite{lenarduzzi2023critical}. By analyzing source code against these rules, PMD generates detailed reports highlighting areas for improvement and enables users to create custom rules for specific analyses \cite{oskouei2018comparing}.

Checkstyle is another static code analysis tool for Java that offers predefined style configurations for standard checks, including Google Java Style and Sun Java Style. Its rules cover various aspects such as annotations, class design, coding, and naming conventions. Checkstyle also supports custom configuration files tailored to user needs~\cite{hovemeyer2004finding, balachandran2013reducing, oskouei2018comparing, lenarduzzi2023critical}.

\begin{figure}[!htbp]
\begin{subfigure}{1\linewidth}
  \centering
  \includegraphics[width=1\linewidth]{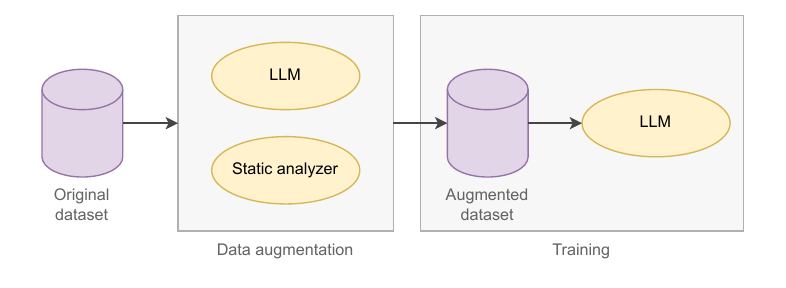}
  \caption{Data augmented training}
  \label{fig:approach1}
\end{subfigure}
\\ \vspace{10pt}
\begin{subfigure}{1\linewidth}
  \centering
  \includegraphics[width=1\linewidth]{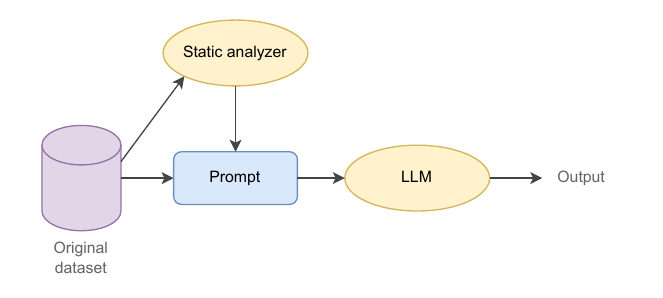}
  \caption{Retrieval augmented generation}
  \label{fig:approach2}
\end{subfigure}
\\ \vspace{10pt}
\begin{subfigure}{1\linewidth}
  \centering
  \includegraphics[width=1\linewidth]{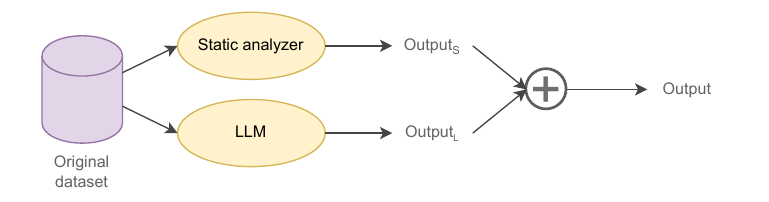}
  \caption{Naive concatenation of outputs}
  \label{fig:approach3}
\end{subfigure}
\caption{Proposed strategies to combine LBS and KBS}
\label{fig:approaches}
\end{figure}

\subsection{Data Augmented Training}

As shown in Figure~\ref{fig:approach1}, this strategy involves retraining the LBS using an augmented dataset \({Da}\), which includes review comments generated by both, static analyzers and the fine-tuned model \(\mathcal{M}_i\). Through this retraining process, the LBS learns from both data sources, producing a more refined model referred to as \(\mathcal{M}_{FT}\).

A simple approach to augmenting the dataset would have been to apply static analysis to the code in \(\mathcal{D}_{\mathcal{M}_i}\) and add or concatenate the generated comments with the existing ones. However, this method does not guarantee data quality within the augmented dataset and fails to account for the insights inferred by the LBS \(\mathcal{M}_i\). 
Therefore, we employ an ensemble learning approach where the two distinct sources—the LBS and the KBS—serve as \emph{experts} to generate data for fine-tuning a model. The underlying rationale is that both KBS and LBS reviews are inherently synthetic. By combining their outputs, we achieve a more balanced and consistent fine-tuning process.

\begin{figure}[htbp!]
  \centering
  \includegraphics[width=\linewidth]{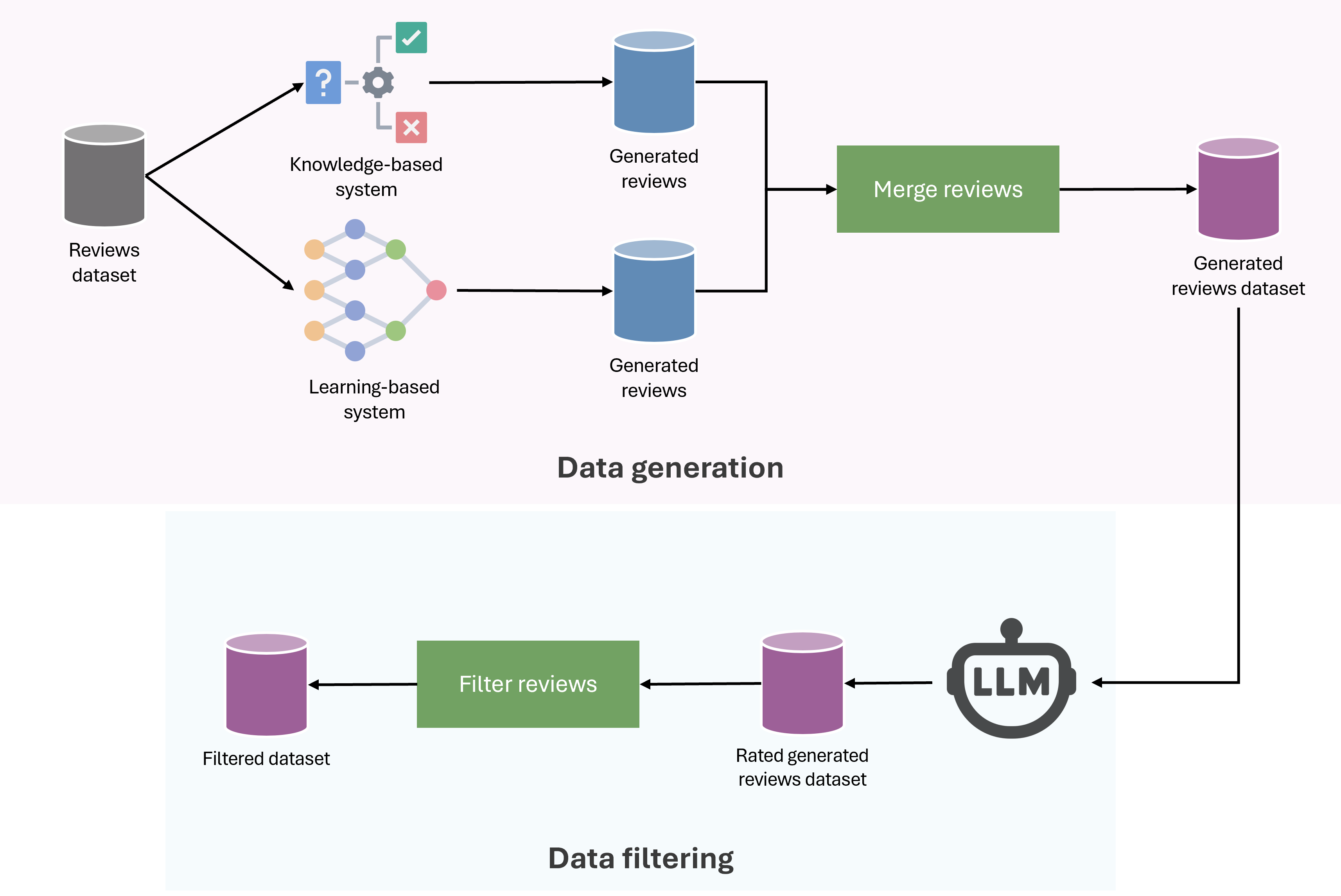}
  \caption{Dataset augmentation pipeline}
  \label{fig:approach} 
\end{figure}

To produce the augmented dataset \(\mathcal{D}_a\), we designed a two-step process (i.e., \emph{data generation} and \emph{data filtering}), as depicted in Figure~\ref{fig:approach}.

In the Data Generation phase, we used the original dataset \(\mathcal{D}_o\) as input. For each code change \( c \), we employed our fine-tuned model \(\mathcal{M}_i\) to generate four context-aware, human-like reviews. Simultaneously, static analyzer rules were applied to each source code \( f \) to produce structured and precise feedback. Each static analyzer generated a report containing several reviews, including the start and end sections of code where each issue was identified. Since our approach focuses on code changes, we extracted the code section highlighted in each review, adding a few context lines before and after each extracted segment.
We then merged the reviews generated by both the static analyzers and \(\mathcal{M}_i\) into a single, unified dataset. Each data point in the dataset consists of tuples in the form \( (f, c, r, t) \), where \( f \) represents the source code file, \( c \) is the code change, \( r \) denotes the review comment, and \( t \) indicates the method used to generate the review (either KBS or LBS). 

After data generation, we applied a systematic Data Filtering step to evaluate and refine the merged dataset, ensuring that only the most relevant and meaningful reviews were retained for each source code. While the fine-tuned model \(\mathcal{M}_i\) can generate context-aware, human-like reviews, its output may sometimes include irrelevant or less meaningful feedback, particularly when handling complex or ambiguous code changes. Similarly, static analyzers, although reliable, may produce output overloaded with false positives \cite{johnson2013don, aniche2020effectiveness}, making it difficult to separate significant concerns from noise.
Therefore, filtering both \(\mathcal{M}_i\)'s and the static analyzers' reviews was essential to maintain a dataset of high-quality, meaningful feedback. This filtering process involved rating each review based on its relevance to the corresponding code. The ratings provided a quantitative measure of the review quality generated by both static analyzers and \(\mathcal{M}_i\).

\begin{figure}[hbt!]
  \centering
  \includegraphics[width=1\linewidth]{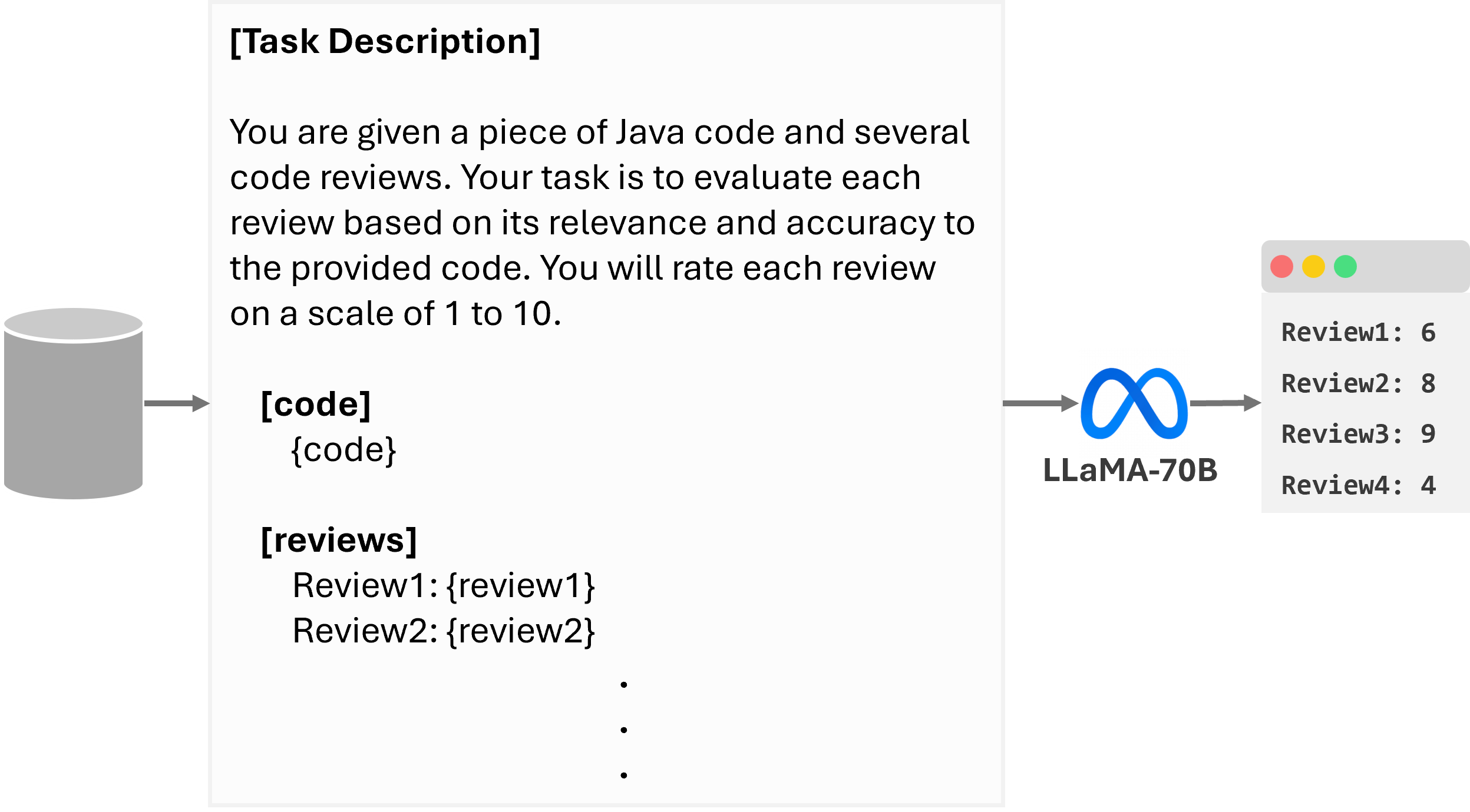}
  \caption{Judgment of review comments using \emph{Llama3-70B}}
  \label{fig:llama3_rating}
\end{figure}

To ensure a fair and scalable rating system, we leveraged large language models, which have demonstrated remarkable performance in similar assessment tasks \cite{zheng2024judging, huang2024empirical, weyssow2024codeultrafeedback}. By using these models, we achieved a more accurate and consistent evaluation of each review, enhancing the dataset's quality and making it a valuable resource for fine-tuning. As shown in Figure~\ref{fig:llama3_rating}, we used \emph{Llama3-70b}, inputting the code and its corresponding reviews. We instructed the model to rate each review on a 10-point scale. A threshold rating of 8 was set, with only reviews surpassing this threshold retained in the final dataset.
After filtering for relevance and quality, we ensured that each comment exceeding the threshold was treated separately. For a source code \(f\) with \(n\) reviews \((r_1...r_n)\), we generated \(n\) distinct data points: \(<f,r_1>\),  \(<f,r_2>\),..., \(<f,r_n>\). Additionally, for each comment, we extracted and included the specific segment of code change related to the issue being addressed, the dataset was then structered as \(<c_1,r_1>\),  \(<c_2,r_2>\),..., \(<c_n,r_n>\).

To prevent overrepresentation of specific rules, we randomly discarded reviews associated with rules that have an excessively high number of reviews. Furthermore, to maintain a balanced dataset, we randomly discarded a subset of learning-based reviews, ensuring an equal distribution between knowledge-based and learning-based reviews.
 
\begin{figure}[!htbp]
  \centering
  \vspace{-2em}
  \includegraphics[width=0.6\linewidth]{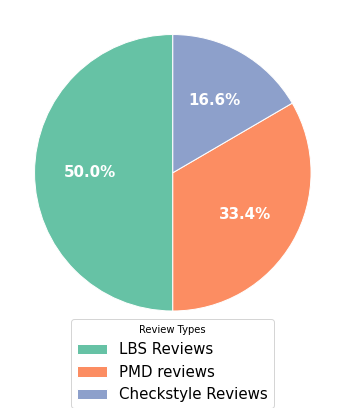}
  \caption{Distribution of LBS and KBS Reviews in Our Dataset}
  \label{fig:chart} 
\end{figure}

The final dataset \(\mathcal{D}_a\) consists of $78,776$ samples, ensuring an equal representation of reviews generated by both KBS and LBS methodologies, as shown in Figure~\ref{fig:chart}. It also ensures a balanced distribution across all KBS rules. 

To obtain the \(\mathcal{M}_{FT}\) model, we partitioned \(\mathcal{D}_a\) dataset into three subsets: 80\% of the samples were assigned to the training set, while the remaining 20\% was equally divided, with 10\% designated for validation and 10\% for testing. Each subset maintained a balanced mix of LBS and KBS reviews. We then fine-tuned the CodeLlama-7b model on this dataset with QLoRA to optimize memory efficiency \cite{hu2021lora}.

\subsection{Retrieval Augmented Generation}

Retrieval-Augmented Generation (RAG) is a technique designed to enhance the generative capabilities of language models by incorporating external knowledge into their prompts during the inference phase \cite{jiang2023active}. In the context of code review, this strategy can be used to embed KBS-generated feedback directly into the prompts of a language-based system, as shown in Figure~\ref{fig:approach2}.

\begin{figure}[htbp!]
  \centering
  \includegraphics[width=0.9\linewidth]{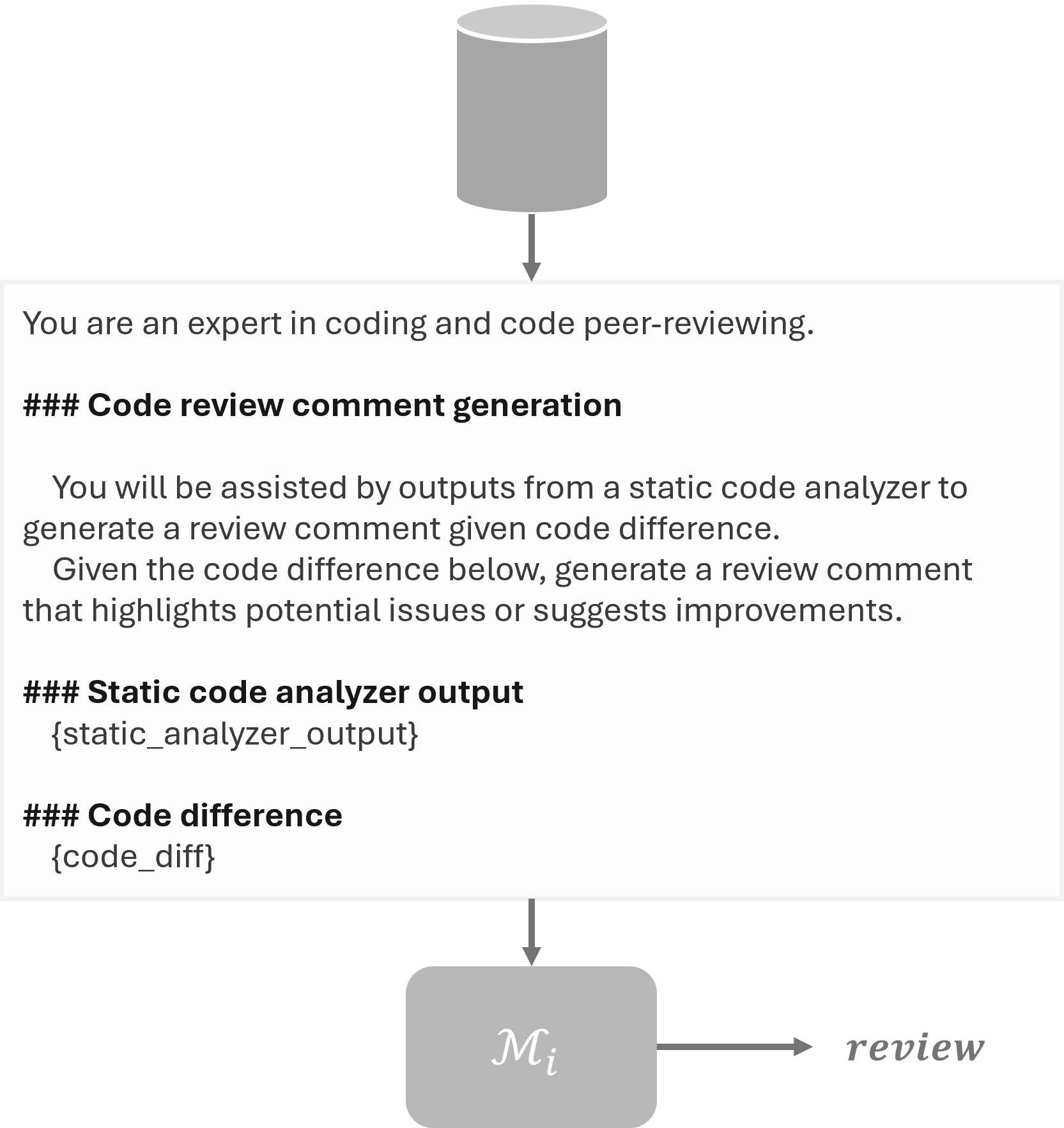}
  \caption{Prompt used to generate review comments using RAG}
  \label{fig:rag_prompt}
\end{figure}

In our approach, the fine-tuned model \(\mathcal{M}_i\) takes as input the code changes from the \(\mathcal{D}_a\) dataset, along with outputs from PMD and Checkstyle. Incorporating KBS knowledge into the prompt guides the model to produce more relevant and precise reviews. This combination ensures that the generated review comments are both comprehensive and contextually informed. As a result, the reviews generated align closely with established coding standards and best practices, thereby enhancing their overall quality. The augmented prompt is illustrated in Figure~\ref{fig:rag_prompt}.

\subsection{Naive Concatenation of Outputs}

The Naive Concatenation of Outputs strategy serves as a baseline approach in which review comments generated separately by the LBS and KBS for the same code are combined to form a single review. As illustrated in Figure~\ref{fig:approach3}, the review comment generated by \(\mathcal{M}_i\) is directly concatenated with the output from the static analyzer (either PMD or Checkstyle).
This approach is straightforward, requiring minimal adjustments to the inference pipeline while ensuring that the final review delivers comprehensive feedback from both systems.

\section{Evaluation}
\label{sec:evaluation}

In this section, we first define the research questions that guide our study, followed by a detailed description of the experimental setup. We then present and interpret the results.
The replication package and the data of our experiments are available online \cite{github_replication, zenodo_14061110}.

\subsection{Research Questions}
The goal of our evaluation is to determine whether integrating knowledge into the LBS enhances the accuracy of generated comments while maintaining adequate coverage compared to using only the LBS or KBS. To structure our evaluation, we define the following research questions:

\textbf{RQ1: Manual evaluation of accuracy:} 
\emph{How accurate are the reviews generated by our hybrid approaches compared to those produced by the baseline?}

To address this question, we perform a manual assessment involving two reviewers on a sample of 10\% of the test set, comparing the accuracy of comments generated by our hybrid approaches against those produced by a fine-tuned language model and static analyzers.

\textbf{RQ2: Alignment of LLM judgments with human assessments (sanity check):}
\emph{To what extent do LLM judgments align with human assessments?}

Given the size of the test set and the five different comment generation options to evaluate, human assessment is impractical for the entire dataset. As an alternative, we employ an LLM to perform assessments across the whole test set. To evaluate the LLM's ability to mimic human judgment, we replicate the manual assessment procedure used for RQ1, substituting the human reviewers with the LLM and measuring the level of agreement between LLM and human evaluations.

\textbf{RQ3: LLM evaluation of generated reviews:}
\emph{How does the accuracy of the reviews generated by our hybrid approaches compare to those produced by the baseline when evaluated by an LLM?}

We use the LLM-as-a-judge approach to assess the accuracy of the generated reviews, comparing results from our hybrid approaches to those from the baseline, specifically the learning-based models and static analyzers.

\textbf{RQ4: Ranking of reviews based on coverage:}
\emph{How do our hybrid approaches compare to the baseline in terms of  coverage?}

For this question, we use an LLM-as-a-judge to conduct comparative evaluations, ranking reviews generated by our approaches and the baseline according to coverage criteria.

\subsection{Experimental Setup}
\begin{figure}[hbt!]
  \centering
  \includegraphics[width=\linewidth]{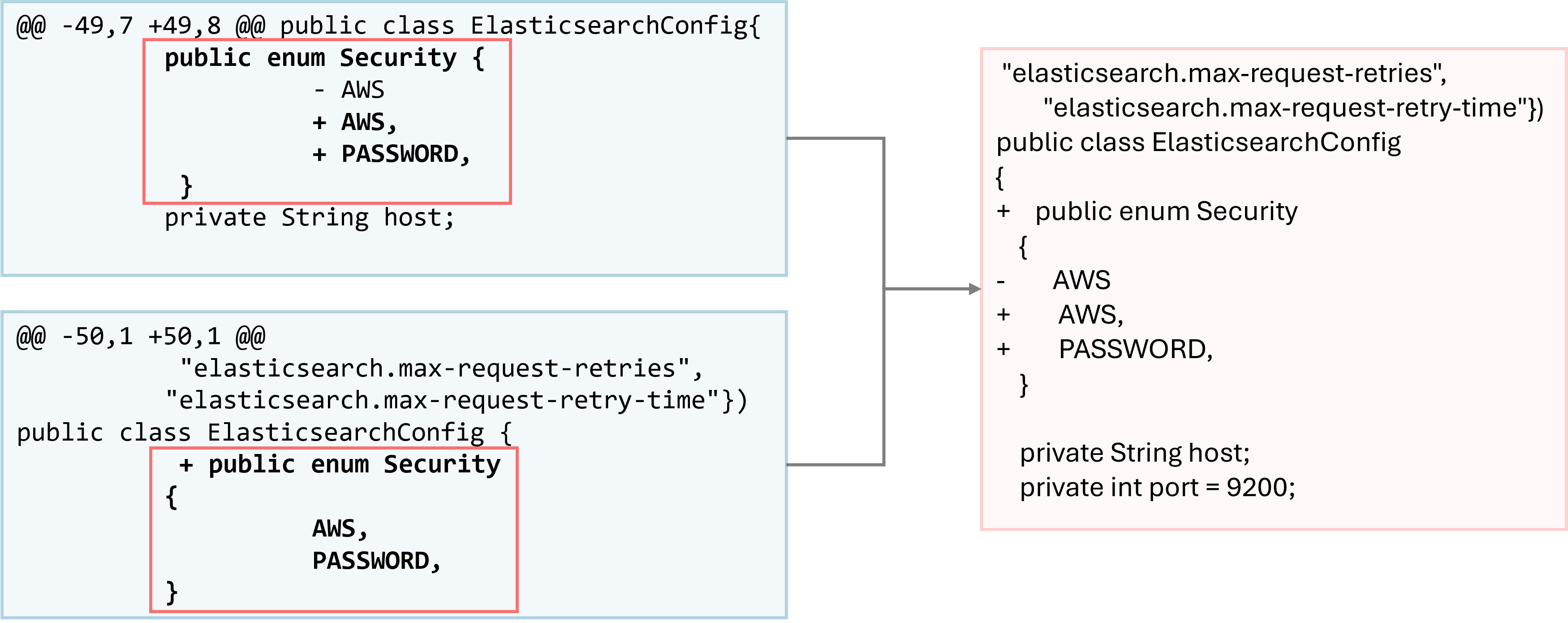}
  \caption{Merging LBS and KBS Changes into a Union Diff}
  \label{fig:overlap} 
\end{figure}

To address our research questions, we conducted a series of experiments. To enable comparison across the different hybrid approaches, we used the test set from our augmented dataset, \(Da\), ensuring that all entries were unseen by the fine-tuned model \(\mathcal{M}_{FT}\). 
Since the retrieval-augmented generation and naive concatenation of outputs approaches are applicable only to code changes with both KBS and LBS reviews, we filtered the test set to include only samples containing reviews from both sources. As illustrated in Figure~\ref{fig:overlap}, we examined the code changes in the selected test set, creating a unified code difference for analysis whenever overlaps were found. For example, if the test set contains a code difference $diff_1$ with an LBS review comment $r_1$ and a code difference $diff_2$ having a KBS review comment $r_2$, we merged both code differences into a unified difference, $diff_u$, as shown in Figure~\ref{fig:overlap}, forming a new triplet ($diff_u$, $r_1$, $r_2$) in our test set. Out of the $7,884$ total samples, we identified $1,245$ common code differences that included both KBS and LBS reviews.

For each of these selected code changes, we generated five types of reviews: the static analyzer review, the review generated by \(\mathcal{M}_i\), the review generated by \(\mathcal{M}_{FT}\), the retrieval-augmented generation review and the naive concatenated review.

For RQ1, we randomly selected $10\%$ of the $1,245$ test set examples for a manual evaluation focused on accuracy. All reviews were anonymized and presented to the evaluators in a randomized order. This approach minimizes confirmation bias, preventing evaluators from unconsciously favoring certain methods by being unaware of the source of each review.

Two evaluators, each with good expertise in code review, assessed each example by classifying reviews as \emph{accurate}, \emph{partially accurate}, or \emph{not accurate}. A review was considered accurate if it correctly addresses issues without any errors or irrelevant information. If only some parts of the review are valid, and the rest are irrelevant or incorrect, it was deemed partially accurate. Finally, a review was classified as not accurate if it fails to address the identified issues and is completely irrelevant to the context. After the initial assessments, a few conflicts were identified and resolved through discussion, leading to a consensus on the final evaluations.

For RQ2, we conducted a sanity check to evaluate the ability of LLMs, specifically \emph{Llama3-70B}, to assess review comments reliably and accurately. Using the same 10\% subset from RQ1, we tasked \emph{Llama3-70B} with categorizing each review as \emph{accurate}, \emph{partially accurate}, or \emph{not accurate}. We then measured the agreement between the ratings of human evaluators and those of \emph{Llama3-70B} by calculating Cohen’s kappa, a statistical measure that quantifies inter-rater reliability for categorical data \cite{thakur2024judging, mchugh2012interrater}. This analysis provides insight into the alignment between LLM and human judgments, helping validate the reliability of LLMs in accurately assessing review comments.

To address RQ3, we instructed \emph{Llama3-70B} to evaluate the code changes across the full dataset of $1,245$ samples, categorizing each review comment into one of three categories: \emph{accurate}, \emph{partially accurate}, or \emph{not accurate}.

For RQ4, We conducted a comparative evaluation by ranking the generated reviews based on their coverage, which measures how effectively the smaller, fine-tuned LLM detects a wide range of code issues. To assess this, we used the larger LLM, \emph{Llama3-70B}, as a reference point, assuming it can identify a comprehensive set of issues. The objective was to evaluate how well the smaller model aligns with this benchmark by capturing as many of these issues as possible. The ranking process reflects the relative effectiveness of each model in identifying coding issues, with Rank 1 assigned to the most comprehensive reviews and Rank 5 to the least.

We then analyzed the \emph{win}-\emph{tie}-\emph{loss} ratios to compare the performance of our three proposed strategies (\emph{DAT}, \emph{RAG}, \emph{NCO}) against the baseline models, $\mathcal{M}_i$ and the static analyzer. A \emph{win} was recorded if a review was ranked at least two levels higher than the baseline, indicating substantially superior coverage, while a \emph{loss} was recorded when a review ranked at least two levels lower, reflecting significantly weaker coverage. Rankings within ±$1$ level were classified as a \emph{tie}, as the difference in coverage was considered minor. This approach ensured a more meaningful evaluation by emphasizing significant differences in issue coverage rather than minor variations.

\subsection{Results}

\subsubsection{Results for RQ1 - Manual evaluation of reviews' accuracy}

\begin{figure}[htbp!]
  \centering
  \includegraphics[width=\linewidth]{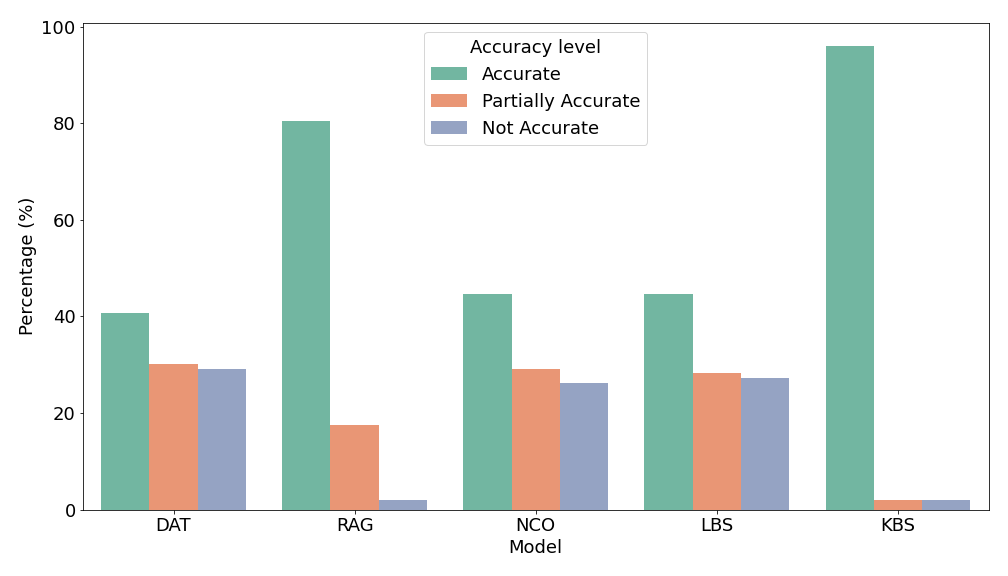}
  \caption{Accuracy levels for the different models based on human assessments}
  \label{fig:manual_judge} 
\end{figure}

The manual evaluation results in Figure~\ref{fig:manual_judge} reveal accuracy differences among the various code review generation approaches. As expected, static analyzers (KBS) consistently demonstrate reliability in producing accurate reviews, whereas the LBS model \(\mathcal{M}_i\) yields less accurate reviews, leading to a higher proportion of partially accurate or inaccurate feedback.
Our hybrid approaches exhibit diverse accuracy levels compared to the baseline systems (i.e., KBS and LBS). Notably, the RAG approach significantly outperforms \(\mathcal{M}_i\) in terms of accuracy. Meanwhile, the fine-tuned model \(\mathcal{M}_{FT}\) produced through the data-augmented training (DAT) approach, and the concatenation approach (NCO) achieve accuracy levels similar to that of \(\mathcal{M}_i\).

In response to \textbf{RQ1}, we conclude that an LLM utilizing the RAG approach generates comments with notably higher accuracy than when used alone. This improvement in accuracy is substantial, although it still falls short of the accuracy achieved by static analyzers. We did not observe an improvement in accuracy for the two other combination approaches.

\subsubsection{Results for RQ2 - Alignment between manual and LLM evaluations}

The alignment between the ratings of human evaluators and those provided by \emph{Llama3-70B} was measured using Cohen's kappa, yielding a score of \textbf{0.72}. This score indicates substantial agreement, highlighting the LLM’s capability to accurately categorize reviews as "accurate," "partially accurate," or "not accurate." Such a high level of concordance suggests that \emph{Llama3-70B} closely mirrors human evaluations in assessing the quality of code reviews, reinforcing its reliability as a substitute for human judgment. Moreover, this degree of agreement aligns with findings from similar studies, further validating its effectiveness as an evaluator.

To answer \textbf{RQ2}, this substantial alignment between \emph{Llama3-70B} and human evaluations enables us to confidently employ the LLM as the evaluator for the entire test set, assessing both accuracy and coverage. Given the observed agreement, we trust that the model will deliver consistent and reliable evaluations across a broader dataset, allowing for efficient and scalable assessment without compromising on quality.

\subsubsection{Results for RQ3 -  LLM evaluation of reviews' accuracy}

\begin{figure}[htbp!]
  \centering
  \includegraphics[width=\linewidth]{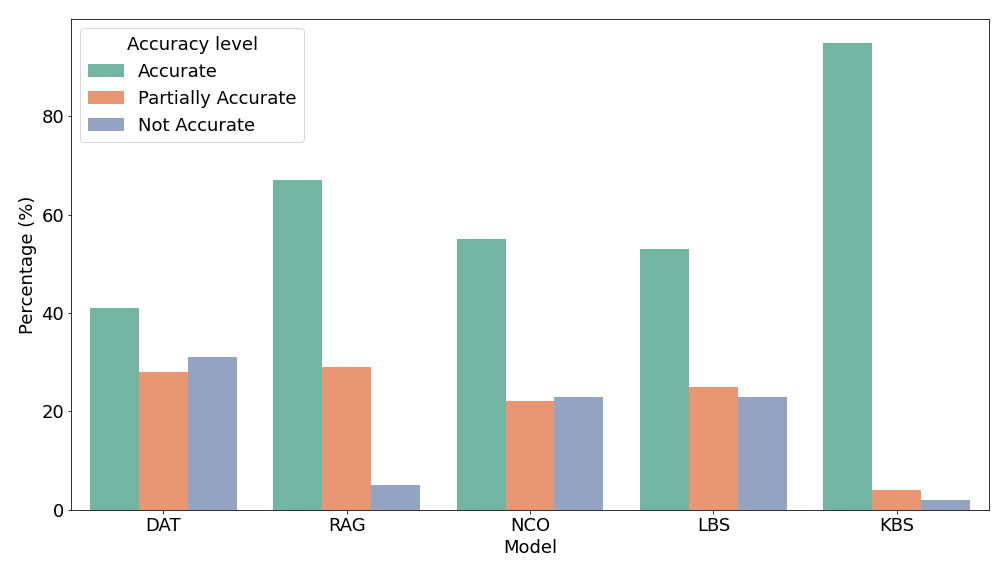}
  \caption{Accuracy levels for the different models based on LLM judgments}
  \label{fig:llm_judge} 
\end{figure}

We employed \emph{Llama3-70B} as an evaluator to assess the review comments generated by each of the five approaches on the test set comprising $1,245$ samples. Each sample included a code change along with five review comments, one from each approach. Figure~\ref{fig:llm_judge} presents the results of this evaluation.
The findings are consistent with those from the manual evaluation in RQ1. As expected, static analyzers (KBS) maintain the highest percentage of accurate reviews due to their deterministic nature in identifying recurring and straightforward coding issues. Similarly, the Retrieval-Augmented Generation (RAG) approach outperforms the original model \(\mathcal{M}_i\) although the improvement in accuracy is slightly less pronounced than what was observed in the smaller RQ1 sample.

The Naive Concatenation of Outputs (NCO) approach achieves comparable accuracy to \(\mathcal{M}_i\), demonstrating its capability to produce reliable reviews. In contrast, the Data-Augmented Training (DAT) approach yielded the lowest accuracy scores among the methods, but still the same levels of NCO and \(\mathcal{M}_i\).

In conclusion, the answer to \textbf{RQ3} is that the RAG approach offers the most significant accuracy improvement over using the LLM alone. The other combination methods, NCO and DAT, show accuracy levels comparable to the standalone LLM model, without notable gains.

\subsubsection{Results for RQ4 - Ranking of Reviews Based on Coverage Criteria Using LLM}
\begin{figure}[htbp!]
  \centering
  \includegraphics[width=\linewidth]{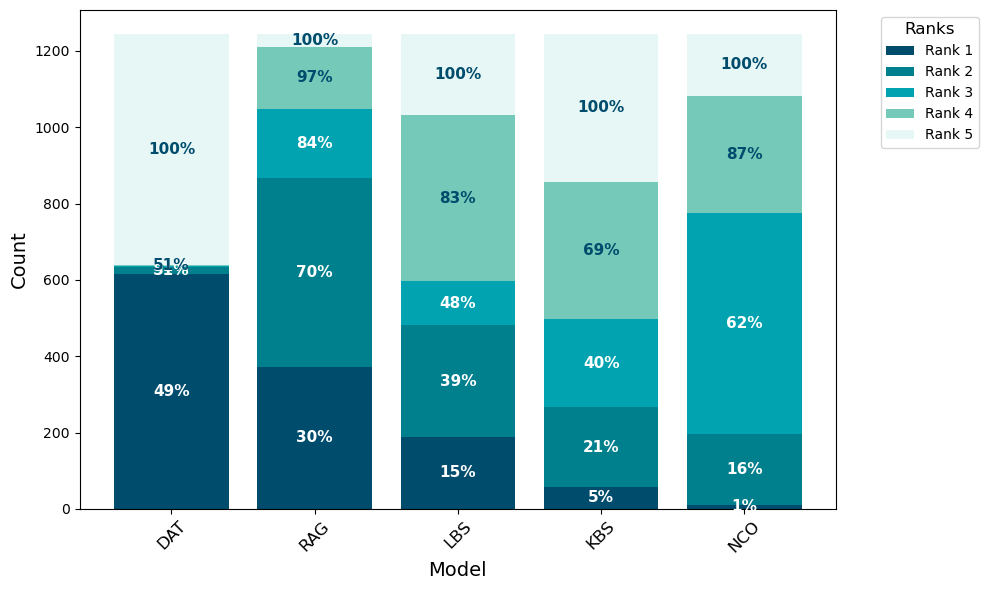}
  \caption{Distribution of Ranks Across Models}
  \label{fig:rating_results} 
\end{figure}

Figure~\ref{fig:rating_results} presents the cumulative ranking distribution for the five code review generation approaches: the baseline methods (KBS and LBS) and the combined approaches (DAT, RAG, and NCO). Rankings assess each approach's effectiveness in achieving comprehensive issue coverage, with Rank 1 indicating the highest coverage and Rank 5 the least.

For the baseline methods, the distribution aligns with expectations. The LBS model displays a broad spread across ranks, with the highest concentration in Rank 4 and less than half of the samples in Ranks 1, 2, and 3. This pattern reflects LBS's real-world training data, likely capturing the selective coverage patterns typical in code review datasets. As a result, LBS effectively identifies common issues but lacks the systematic, rule-based coverage characteristic of KBS. Static analyzers, in contrast, are predominantly ranked in Ranks 4 and 5 due to their rule-based approach, which, while precise, lacks adaptability and the capacity to detect a broader range of issues. This rule-bound structure restricts static analyzers to specific, predictable issue types, limiting their ability to provide comprehensive coverage as LBS might.

The DAT approach has a high proportion of reviews achieving Rank 1 (49\%), demonstrating strong coverage in these cases. Notably, the DAT distribution is bimodal, with relatively few reviews in Ranks 2, 3, or 4. This suggests that DAT either ranks at the top, addressing the majority of issues in a code change, or falls to Rank 5 when its coverage is limited. This polarized performance pattern in \(\mathcal{M}_{FT}\) after data augmentation shows that it either achieves comprehensive coverage or misses key areas, rarely achieving intermediate coverage levels.
RAG performs consistently well, with most reviews in Ranks 1 and 2 (70\%). This suggests that RAG reliably covers a substantial number of issues in code changes, with a notable share in Rank 2, indicating that it frequently provides strong, if not complete, coverage.

The NCO approach primarily appears in middle ranks, especially in Rank 3, indicating moderate coverage. This outcome reflects the straightforward concatenation process, which combines reviews from LBS and KBS without advanced integration. As a result, NCO's coverage is influenced by the quality of the LBS review; when LBS has low coverage, NCO’s coverage also tends to be limited. However, the addition of static analyzer reviews offers slight improvements, giving NCO somewhat broader coverage than LBS alone in the top three ranks.

In summary, DAT and RAG emerge as the top-performing approaches, together accounting for nearly 80\% of Rank 1 reviews, each excelling in one of the top two ranks. NCO demonstrates moderate performance, with Rank 3 as its most common position.

\begin{figure*}[htbp!]
  \centering
  \includegraphics[width=1\linewidth]{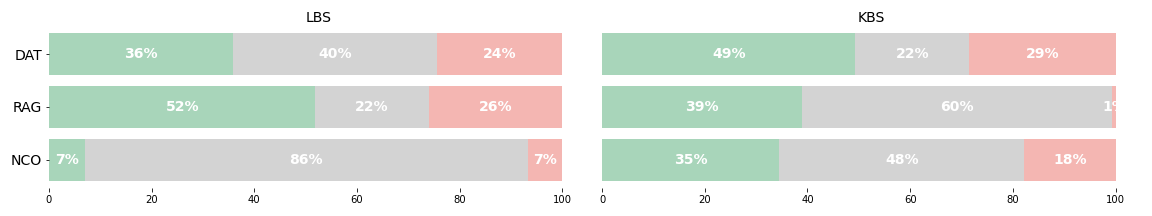}
  \caption{Win-tie-loss ratios of our proposed approaches compared to the baseline in terms of coverage. }
  \label{fig:wtl}
\end{figure*}

To enable a direct comparison between the combination approaches and baselines, we conducted a pairwise win-tie-loss analysis based on rankings. As shown in Figure~\ref{fig:wtl}, DAT has a high win rate against both LBS and KBS, highlighting its effectiveness in achieving broader coverage. RAG also performs well, consistently identifying a wide range of issues and frequently surpassing LBS in coverage. In contrast, while NCO positions favorably against KBS, it shows a higher tie rate with LBS, indicating that it generally achieves moderate coverage but does not consistently outperform LBS. Cases where NCO provides less coverage than KBS can arise when LBS-generated reviews contradict or override KBS output, reducing the overall coverage of the concatenated result. For example, in one instance, the KBS produced the following comment:
"Unused import 'com.google.common.collect.ImmutableList.'"
However, the LBS-generated review effectively invalidated this observation:
"The code change adds a new import statement for com.google.common.collect.ImmutableList. This is a good addition, as it allows the class to use the ImmutableList class, which can be useful for creating immutable lists."
In such cases, the conflicting feedback from LBS can diminish the impact of KBS findings, leading to lower coverage in the NCO approach.

To answer RQ4, we conclude that two of the combination approaches, DAT and RAG, significantly enhance the coverage of issues that can be automatically identified compared to the baselines.

\subsection{Discussion}
Based on the results from RQ3 and RQ4, our baseline models performed as expected, each with distinct strengths and limitations. The knowledge-based system (KBS) demonstrated high accuracy in detecting rule-based issues, particularly for violations that could be systematically encoded. 
However, its coverage was limited, missing many complex issues that require a nuanced understanding of context, which cannot easily be captured through static rules alone. 

In cases where the KBS was deemed inaccurate, its reliance on predefined rules is often the primary cause, as these rules may generate irrelevant or outdated reviews. Since they do not always align with the specific context of the code being reviewed, they can lead to misjudgments. For example, rules based on fixed thresholds (e.g., maximum method length or parameter list size) may result in incorrect assessments, particularly when they fail to account for context-specific factors.

For example, as seen in Table~\ref{tab:exp}, KBS generated a review comment addressing a basic violation of naming conventions, which is straightforward but ranks lowest. In contrast, the learning-based system \(\mathcal{M}_i\), ranked fourth, identified a more subtle issue that would be challenging to encode in a rule-based analyzer, highlighting the advantage of LBS in identifying context-dependent issues.

The Data Augmented Training (DAT) approach expanded the coverage beyond \(M_i\) by exposing the model to a wider variety of issues, as the augmented dataset \(Da\) allowed the model to handle both LBS and KBS issues more effectively. This exposure enabled DAT to consider contextual factors and recognize project-specific nuances that KBS would typically overlook, while also incorporating rule-based issues where appropriate. For example, in Table~\ref{tab:exp}, DAT achieved the highest rank by generating a detailed and context-sensitive review comment that covered both rule-based and nuanced observations, reflecting its expanded coverage capabilities.

The Retrieval-Augmented Generation (RAG) approach enhanced both the accuracy and coverage of generated reviews by incorporating information retrieved from KBS. This retrieval step enriched RAG's feedback, making it more comprehensive and relevant. However, it still fell short of KBS’s precision on rule-based issues due to \(\mathcal{M}_i\)’s inherent inaccuracies during inference. In Table~\ref{tab:exp}, RAG’s comment, ranked second, provides a more detailed critique than the baseline models, yet it lacks the depth of DAT’s analysis.

The Naive Concatenation of Outputs (NCO) strategy did not notably improve accuracy and offered only a marginal increase in coverage. Since part of the NCO approach’s output is directly inherited from the LBS, its accuracy is inherently tied to the quality of the LBS-generated comments. If the LBS produces a partially accurate comment, the corresponding NCO output will also reflect this partial accuracy, as the NCO does not refine or correct LBS-generated reviews.

Table~\ref{tab:exp} illustrates an NCO’s generated comment, ranked third, integrated feedback from both baselines, but without advanced synthesis, resulting in a moderate improvement. Although NCO included the LBS and KBS comments, it was unable to address the more complex issues that DAT and RAG identified.

In summary, DAT and RAG stood out for their enhanced coverage, capturing nearly 80\% of Rank 1 assessments. NCO, by contrast, demonstrated only moderate performance, frequently achieving Rank 3 due to its limited integration of baseline insights. This analysis underscores the strengths of hybrid approaches in combining rule-based precision with learning-based adaptability to achieve comprehensive code review coverage.

\begin{table*}[!htbp]
\centering
\caption{Comparison of Review Comments. Excerpts of comments are shown for brevity.}
\label{tab:exp}
\begin{tabularx}{1\linewidth}{X}
\toprule
\multicolumn{1}{c}{\textbf{Code change}} \\
\midrule
@@ -38,5 +38,7 @@ public class GoogleAuthenticatorAccount extends OneTimeTokenAccount \{

                            @JsonProperty("validationCode") final int validationCode,
                            
                            @JsonProperty("scratchCodes") final List$\langle Integer\rangle$
                            scratchCodes) \{
                            
         super(username, secretKey, validationCode, scratchCodes);
         
+        long v =  java.lang.System.currentTimeMillis();

+        this.setId(v);
     \}
 \}
\\ \midrule
\multicolumn{1}{c}{\textbf{DAT}} \\ \midrule
\#\#\#\# Potential Issues:

- The setId() method uses `System.currentTimeMillis()` in the constructor, which could lead to issues if the object is not persisted before the constructor finishes.

- The `id` field should be marked as transient to avoid unnecessary persistence.

-  It is possible to remove the final keyword to allow subclasses to override the constructor.

\#\#\#\# Areas for Improvement:

- Consider adding a `toString()` method for better readability.

- Validate the constructor parameters (username, secretKey, etc.) to ensure they are not null or empty.
\\ \midrule
\multicolumn{1}{c}{\textbf{RAG}} \\ \midrule
The use of the short variable name `v` is discouraged. Consider using more descriptive names like `currentTimestamp` or `uniqueIdentifier`. Also, `System.currentTimeMillis()` may not be the most efficient way to generate a unique identifier; alternatives like `UUID.randomUUID()` or `SecureRandom.nextLong()` could be more robust.
\\ \midrule
\multicolumn{1}{c}{\textbf{NCO}} \\ \midrule
The addition of the `id` field is unnecessary as the parent class `OneTimeTokenAccount` already has an `id` field. Using the existing field would avoid redundancy and confusion.

Avoid using short variable names like `v`. 

\\ \midrule
\multicolumn{1}{c}{\textbf{LBS}} \\ \midrule
The addition of the `id` field is unnecessary as the parent class `OneTimeTokenAccount` already has an `id` field. Using the existing field would avoid redundancy and confusion.
\\ \midrule
\multicolumn{1}{c}{\textbf{KBS}} \\ \midrule
Avoid using short variable names like `v`. 
\\ \bottomrule

\end{tabularx}
\end{table*}

\subsection{Threats to Validity}
During our evaluation process, we identified several potential threats to the validity of our results. The most notable threats are discussed below.

Evaluator Bias: While human evaluations were conducted to assess accuracy in RQ1, there is a potential for evaluator bias. To mitigate this threat, we presented the reviews in an anonymized and randomized order to reduce any unconscious preference for particular methods.

Generalizability: Our study primarily focuses on Java code, using tools and datasets specific to the Java programming language. This may limit the generalizability of our findings to other languages and contexts, where static analyzers or learning-based systems could exhibit different strengths and weaknesses.

Dependency on LLM Judgment: For RQ3 and RQ4, we relied on an LLM (\emph{Llama3-70B}) to evaluate accuracy and coverage across the entire test set. Although Cohen’s kappa indicates substantial agreement with human evaluations on a subset of data, subtle variations in the LLM's judgment across the full dataset may not fully align with human assessments, potentially impacting reliability. To address this, in addition to the kappa score, we manually inspected a random subset of evaluations for RQ4 and found them consistent and reliable.

Choice of Static Analyzers: The effectiveness of the KBS approach depends on the static analyzers selected. We chose PMD and Checkstyle, which we believe are state-of-the-art tools; however, other static analyzers might yield different results. Our confidence in these tools is based on their widespread use and established effectiveness in rule-based code analysis.

Evaluation Metrics Limitations: The ranking system (win-tie-loss) used in RQ4 considers the number of levels above or below baseline models, but it may not capture subtle differences in review quality. To account for this, we adopted a conservative approach, requiring at least a two-level ranking difference for a “win,” to ensure that only substantial improvements were counted as wins.

\section{Related work}
\label{sec:literature}

\subsection{Code review automation}
Several approaches were proposed to assist code review. Gupta et al. \cite{gupta2018intelligent} introduced an LSTM-based model trained on positive and negative (code, review) pairs, selecting candidate reviews based on code similarity and predicting relevance scores. Siow et al. \cite{siow2020core} enhanced this with multi-level embeddings, leveraging word-level and character-level representations to better capture the semantics of code and reviews.

With the advent of large language models, the focus has shifted toward generative models to fully automate code review tasks. Tufano et al. \cite{tufan2021towards} developed a transformer-based model to suggest code edits before and after code review, later enhancing it by pre-training T5 on Java and technical English~\cite{tufano2022using}. Li et al. \cite{li2022automating} pre-trained CodeT5 on a multilingual dataset and fine-tuned it for code review tasks like quality estimation, review generation, and code refinement. Sghaier et al. \cite{ben2024improving} further advanced this area by applying cross-task knowledge distillation to address successive code review tasks jointly, enhancing performance and promoting tasks' interdependence.

Current research efforts have significantly advanced the automation of code review, introducing different and innovative approaches that enhance code review tasks. However, the performance of these automated approaches remains limited in terms of correctness, as indicated by low BLEU scores, suggesting that further refinement is needed to achieve higher accuracy and reliability as expected in practical software development contexts.

\subsection{LLMs and static analysis combination}

Recent research has increasingly focused on enhancing LLM-based solutions for software engineering using several techniques. One is by integrating them with static analysis tools, addressing the challenge of reducing inaccurate or incomplete results. 

In automated program repair, RepairAgent \cite{bouzenia2024repairagent} employs static analysis to gather contextual data that guides LLM-driven code correction, while PyTy \cite{chow2024pyty} relies on type-checking mechanisms to validate the accuracy of LLM-generated candidates in resolving static type inconsistencies. For software testing, approaches like TECO \cite{nie2023learning} apply static analysis to derive semantic features for training transformers in test completion, while ChatTester \cite{yuan2023no} and TestPilot \cite{schafer2023adaptive} utilize similar techniques to prepare contextual information that supports iterative LLM-based test code repair processes. For bug detection, LLMs were combined with static analysis to reduce false positives. SkipAnalyzer \cite{mohajer2023skipanalyzer} and GPTScan~\cite{sun2024gptscan} use static analysis to validate LLM predictions, while D2A~\cite{zheng2021d2a} and ReposVul \cite{wang2024reposvul} refine bug labeling and re-rank predictions.
For code completion, STALL+ \cite{liu2024stall+} integrates static analyzers with LLMs through a multi-phase approach involving prompting, decoding, and post-processing.

These studies illustrate the effectiveness of combining LLMs with static analysis across tasks like program repair, bug detection, testing, and code completion. However, this integration has yet to be explored for code review.

\section{Conclusion}
\label{sec:conclusion}

In this paper, we explored hybrid approaches that combine knowledge-based systems (KBS) with large language models (LLMs) to enhance the automation of code review. By integrating KBS-derived insights at three stages—data preparation (Data-Augmented Training, DAT), inference (Retrieval-Augmented Generation, RAG), and post-inference (Naive Concatenation of Outputs, NCO)—we aimed to leverage the strengths of both KBS and learning-based systems (LBS) to generate more accurate and comprehensive code review comments.
Our empirical evaluation showed that combination approaches offer distinct advantages. RAG emerged as the most effective, improving both accuracy and coverage of review comments by enriching the generation process with structured, contextually relevant knowledge from static analysis tools. DAT achieved broad coverage by exposing the LLM to diverse issue types in training, sometimes at the expense of precision. NCO, while straightforward, achieved moderate improvements in coverage. 

These findings underscore the potential of combining static analysis tools with LLMs to address the limitations of individual approaches in automated code review. Future work will involve exploring more sophisticated integration of knowledge into open-weight LLMs. We also plan to expand this methodology to additional programming languages and exploring bytecode-level analysis for greater depth. Furthermore, the integration of more advanced static analyzers and dynamic analysis tools could further enhance coverage and precision, ultimately contributing to a more robust and versatile code review automation pipeline.

\bibliographystyle{IEEEtran}
\bibliography{references}

\end{document}